\documentclass{article}

\usepackage[square, numbers]{natbib}
     \usepackage[final, nonatbib]{neurips_2019}

\usepackage[utf8]{inputenc} 
\usepackage[T1]{fontenc}    
\usepackage{url}            
\usepackage{booktabs}       
\usepackage{amsfonts}       
\usepackage{nicefrac}       
\usepackage{microtype}      
\usepackage{amsfonts}
\usepackage{amssymb}
\usepackage{amsmath}
\usepackage{amsthm}
\usepackage[stable]{footmisc}
\usepackage{graphicx}

\DeclareMathOperator*{\argmin}{arg\,min}

\newcommand{\Pb}{\mathbb{P}}
\newcommand{\Qb}{\mathbb{Q}}

\newcommand{\E}{\mathbb{E}}
\newcommand{\R}{\mathbb{R}}
\newcommand{\dopt}{d^\text{opt}}
\renewcommand{\P}{\mathbb{P}}

\DeclareSymbolFont{bbold}{U}{bbold}{m}{n}
\DeclareSymbolFontAlphabet{\mathbbold}{bbold}
\newcommand{\one}{\mathbbold{1}}
\newcommand\ind{\protect\mathpalette{\protect\independenT}{\perp}}
\def\independenT#1#2{\mathrel{\rlap{$#1#2$}\mkern2mu{#1#2}}}

\title{When the Oracle Misleads: Modeling the Consequences of Using Observable Rather than Potential Outcomes in Risk Assessment Instruments}

\author{%
  Alan Mishler \\
  Department of Statistics \& Data Science\\
  Carnegie Mellon University\\
  Pittsburgh, PA \\
  \texttt{amishler@stat.cmu.edu} \\
  \And
  Niccol\`o Dalmasso \\
  Department of Statistics \& Data Science\\
  Carnegie Mellon University\\
  Pittsburgh, PA \\
  \texttt{ndalmass@stat.cmu.edu}
}

\begin{document}

\maketitle

\begin{abstract}
    Risk Assessment Instruments (RAIs) are widely used to forecast adverse outcomes in domains such as healthcare and criminal justice. RAIs are commonly trained on observational data and are optimized to predict observable outcomes rather than potential outcomes, which are the outcomes that would occur absent a particular intervention. Examples of relevant potential outcomes include whether a patient's condition would worsen without treatment or whether a defendant would recidivate if released pretrial. We illustrate how RAIs which are trained to predict observable outcomes can lead to worse decision making, causing precisely the types of harm they are intended to prevent. This can occur even when the predictors are Bayes-optimal and there is no unmeasured confounding.
\end{abstract}

\section{Introduction\footnote{This paper was presented at the workshop \emph{“Do the right thing”: machine learning and causal inference for improved decision making}, NeurIPS 2019. \protect\url{https://tripods.cis.cornell.edu/neurips19_causalml/}}}
Machine learning is increasingly widely used to support decision making in domains as diverse as healthcare, criminal justice, and consumer finance. In particular, predictive models are often used to estimate the risk of a negative outcome such as death, recidivism, or default on a loan \citep{kourou_machine_2015, Caruana2015Pneumonia, colubri_transforming_2016, brennan_evaluating_2009, khandani_consumer_2010}. Scores from these Risk Assessment Instruments (RAIs) are made available to decision makers, such as doctors, judges, or loan officers, who may take them into account when deciding whether or not to admit a patient to a hospital, release a defendant on bail, or issue a loan to an applicant.

When the decision maker's goal is to reduce the risk of the predicted outcome, they are naturally concerned with \emph{potential outcomes}, the outcomes that would occur under each available decision. When these outcomes correspond to an intervention that actually takes place, they are observable; otherwise, they are counterfactual. (Many authors use ``counterfactual outcomes'' as a synonym for potential outcomes.) RAIs are typically trained on observational data, in which outcomes are affected by historical decisions, and they are typically designed to predict exclusively observable outcomes. Hence, these RAIs can only be sensibly understood as predicting the risk of an outcome \emph{under the historical decision process that generated the data}; they are not generally appropriate for helping decision makers decide among different courses of action.

Although RAIs based on potential outcomes have been proposed in the context of medicine \citep{schulam_reliable_2017, shalit_estimating_2017} and recidivism prediction \citep{mishler_modeling_2019}, RAIs designed to predict observable outcomes are in widespread use. While many of the limitations of such RAIs have been acknowledged \citep{chen_machine_2017, veale_fairness_2018}, and problems associated with particular RAIs have been investigated \citep{povyakalo_how_2013, lum_predict_2016}, there does not appear to be a general mathematical model that provides insight into how and why such RAIs can lead users astray.

In this work, we aim to fill this gap, showing how RAIs based on observable outcomes can lead to \emph{worse} outcomes, i.e., more severe departures from an optimal treatment regime, than before the RAI was introduced. This has nothing to do with the quality of prediction; it can occur even when (1) the oracle predictor is available and (2) there is no unmeasured confounding. We describe several dangerous properties of these RAIs and illustrate their suboptimality with a simple example.

\section{Setup: RAIs and optimal treatment regimes}
We anchor the problem in the context of medicine, but the results generalize to any domain where an estimated risk is used to drive decision making designed to mitigate that risk.

Suppose that at time $t = 0$ we have random variables drawn from a counterfactual distribution $(U, X, A, Y^0, Y^1, Y) \sim \Qb_0$, where $U\in\R^{p'}$ is a set of unobserved confounders, $X\in\R^p$ is a set of observed covariates, $A\in\{0, 1\}$ is a binary treatment or intervention decision, and $Y\in\{0, 1\}$ is an outcome, with $Y = 1$ indicating an adverse event such as patient death. $Y^0$ and $Y^1$ denote the potential outcomes under treatment decisions $A = 0, 1$. Let $\Pb_0$ denote the marginal distribution of the observable vector $(X, A, Y)$ at $t = 0$. We use $\E_t$ and $\Pb_t$ to denote expectations and probabilities at time $t$, but when these do not change over time we drop the subscript and use $\E$ and $\Pb$. 

\begin{figure}
\centering
    \begin{minipage}{0.45\linewidth}
    \centering
    \includegraphics[width=0.35\linewidth]{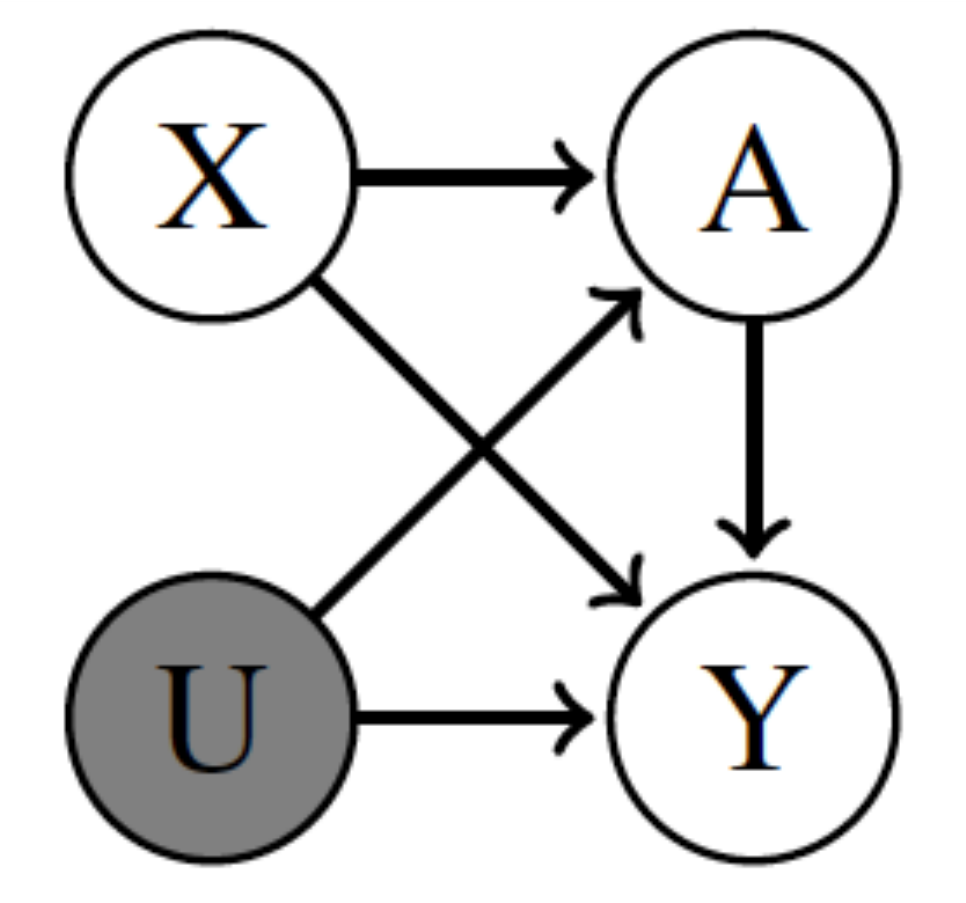}
    \caption{Causal graph at time $t=0$, with unobserved confounders $U$.\label{f:setup_t0}}
    \end{minipage}
    \hspace{0.5cm}
    \begin{minipage}{0.45\linewidth}
    \centering
    \includegraphics[width=0.33\linewidth]{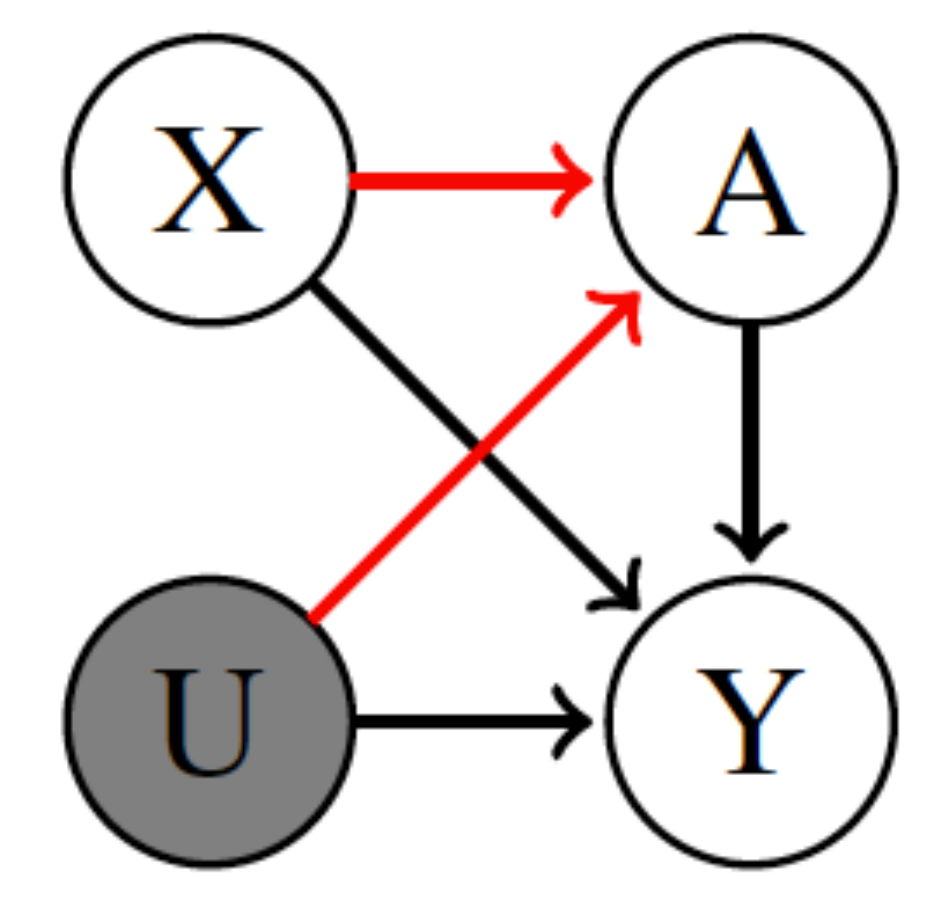}
    \caption{Causal graph at time $t=1$, with possibly changed treatment decision process.\label{f:setup_t1}}
    \end{minipage}    
\end{figure}

Now suppose that iid data drawn from $\Pb_0$ is used to construct a predictor $s(X)$ of $Y$ given $X$. For example, suppose that $s(X) = \hat{\E}_0[Y|X]$. This predictor is made available to decision makers in the form of an RAI, as a ``risk score,'' giving rise at time $t = 1$ to new distributions $(U, X, A, Y^0, Y^1, Y) \sim \Qb_1$ and $(X, A, Y) \sim \Pb_1$. We make the following assumptions at all time points $t$:
\begin{enumerate}
    \item $Y = AY^1 + (1 - A)Y^0$ \emph{(Consistency)} 
    \item $\Pb_t[\Pb_t(0 < \pi_t(X) < 1)] = 1$ \emph{(Positivity)}
    \item $A \ind Y^a|X, U$, for $a \in \{0, 1\}$ \emph{(No confounders beyond $X$ and $U$)}
    \item $(U, X, Y^0, Y^1)_{\Qb_t} \overset{d}{=} (U, X, Y^0, Y^1)_{\Qb_{t+1}}$ \emph{(Only the treatment and outcome change after the RAI is introduced.)}
    \item $0 < \Pb_t(Y^1 < Y^0) < 1$ \emph{(Treatment sometimes helps and sometimes hurts overall. For example, hospitalization can expose patients to dangers such as MRSA or medical errors.)}
\end{enumerate}
Note that $U$ is not observable by the researchers who construct $s(X)$, but it may include variables that are available to doctors at the time they render a treatment decision. That is, the treatment decision process may change in light of the new RAI, but the RAI does not otherwise affect patient outcomes or the distribution of covariates. Causal graphs representing the change in the decision process from time 0 to time 1 are given in Figures \ref{f:setup_t0} and \ref{f:setup_t1}.
 
Given all possible treatment decision functions $\mathcal{D} = \{d: (X, U) \mapsto \{0, 1\}\}$, it is easy to show that the optimal treatment regime with respect to the expectation of $Y$ is
\begin{align} \label{eq:optimal_decision}
    \dopt(X, U) &:= \argmin_{d\in\mathcal{D}} \E[Y^{d(X, U)}] = \one\{\E[Y^1|X, U] < \E[Y^0|X, U]\}.
\end{align}
where the expectations in this expression do not change over time, as a consequence of Assumption 4. Given that $s(X)$ is not designed as an estimator of $\dopt$, the questions of interest are:
\begin{enumerate}
    \item When is $\E_1[Y] \leq \E_0[Y]$, as desired? That is, when does the RAI make things better, or at least not worse?
    \item How far is $\E_1[Y]$ from $\E_1[Y^{\dopt}]$, the optimal outcome?
\end{enumerate}
We are also interested in versions of these questions where the quantities are conditional on $(X, U)$. For example, we wish to know when outcomes get better or worse differentially for patients from different demographic groups, which could cause the RAI to be considered unfair.

\section{RAIs can make things worse}
Let $\pi_t(X) = \Pb_t(A = 1|X, U)$ denote the treatment propensity at time $t$, with $\Gamma(X, U) := \pi_1(X, U) - \pi_0(X, U)$, and let $\mu^a(X, U) = \E[Y|X, U, A = a]$ denote the outcome regression functions, for $a \in \{0, 1\}$. $\mu^a$ does not change over time, per assumptions 3 and 4. We have:
\begin{align}
	\Delta := \E_1[Y] - \E_0[Y]
	                  &= \E\left\{\Gamma(X, U)(\mu^1(X,U) - \mu^0(X,U))\right\} \label{eq:basic}
\end{align}
(See the derivation in the Appendix). It is easy to see that $\Delta$ can be positive, meaning that more patients die after the introduction of the RAI, and that even if it is negative, outcomes could worsen for particular strata of $(X, U)$. For example, consider a subpopulation for whom $\mu^1(X,U) < \mu^0(X,U)$ and $\Gamma(X, U) < 0$. These could be patients who historically benefited from hospitalization and were hospitalized at high rates, so that $\E_0[Y|X]$, their likelihood of death in the training data, is small. The apparent low risk could prompt doctors to reduce the rate at which they hospitalize these patients, causing death rates to rise. Of course, if $\Delta$ is positive, then $\E_1[Y] - \E[Y^{\dopt}]$ will be positive as well.

For ease of exposition, we now restrict our attention to a special case of the above scenario, wherein $U = \emptyset$, so that there is no unmeasured confounding, and $s(X) = \E[Y|X]$, so we have access to the MSE-minimal oracle predictor. We suppose that once the RAI is introduced, decisions are made deterministically according to a threshold rule $d(x) = \one\{s(X) \geq \theta\}$ for some $\theta \in [0, 1]$. That is, doctors hospitalize patients iff their estimated risk is at or above $\theta$. We illustrate with a toy example.

\subsection{Toy example}
We assume a single covariate $X \sim \text{Unif}(0, 1)$, representing a marker of disease severity. We let both the treatment propensity and the risk of non-treatment increase in $X$, with $\pi_0(X) = \E[Y^0|X] = X$, and we let the risk of treatment be $\E[Y^1|X] = (0.7-X)^2$. This represents a situation in which treatment is beneficial on average above a certain level of $X$ but harmful otherwise.

\begin{figure}
    \centering
    \includegraphics[width=1.0\linewidth]{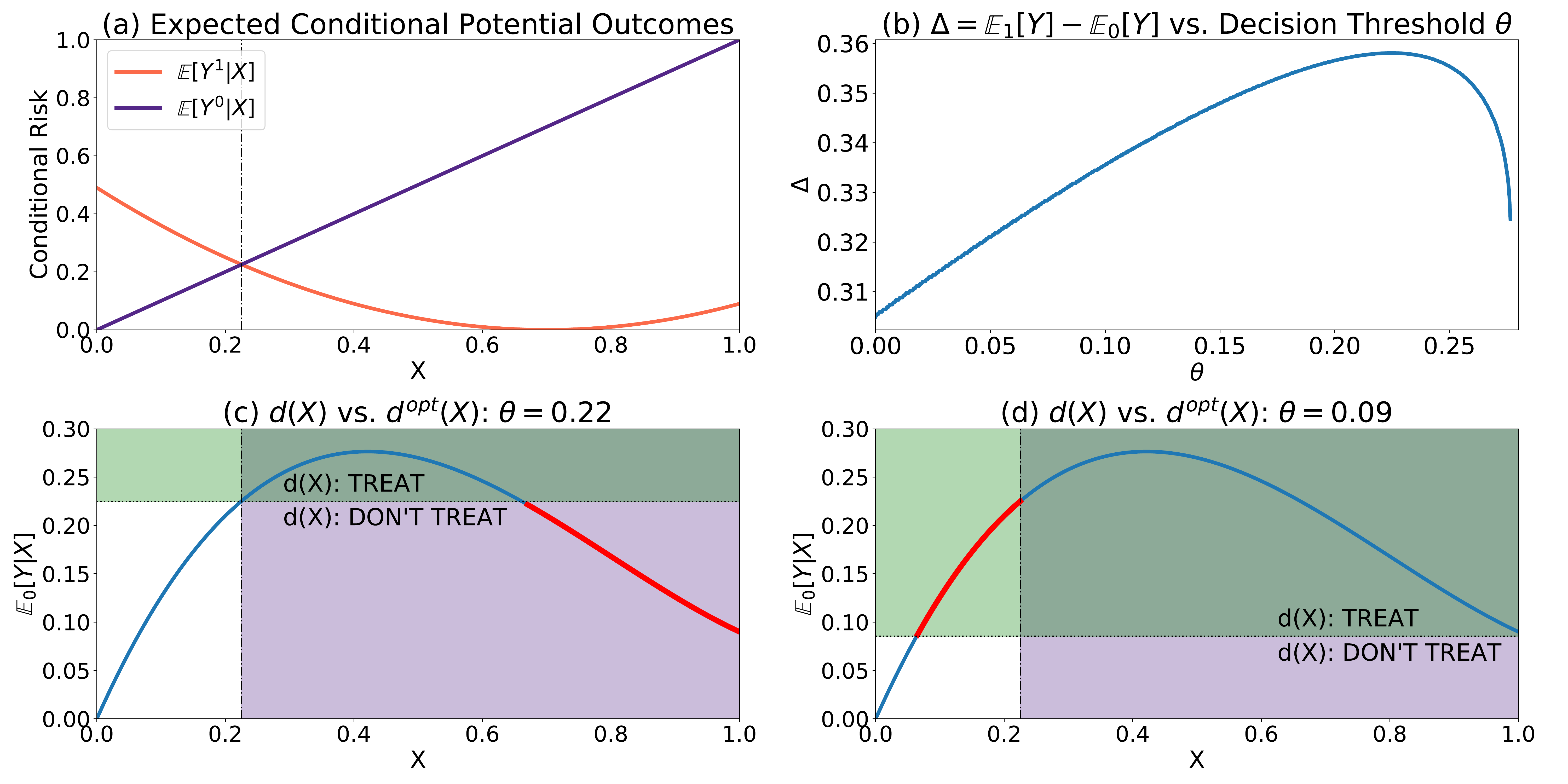}
    \caption{\small{(a) Conditional expectations in the toy example. (b) Behavior of $\Delta=\mathbb{E}_1[Y] - \mathbb{E}_0[Y]$ as a function of the cutoff $\theta$. (c) and (d) show groups treated at time 1 under $d(X)$ and $\dopt(X)$, for two possible values of $\theta$. The optimal treatment group is $\{X > 0.22\}$, in purple. The group treated under $d(X)$ is $\{\E_0[Y|X] \geq \theta\}$, in green. Red lines indicate groups that are harmed under $d(X)$ as a result of not receiving or receiving treatment, respectively.}}
    \label{fig:toy_example}
\end{figure}

Figure \ref{fig:toy_example} (a) shows the two conditional expectations $\E[Y^0|X], \E[Y^1|X]$. The optimal treatment rule here is $\dopt(X) = \one(X \geq 0.22)$, indicated by the dashed line. The rule that is actually implemented at time $t = 1$ is $d(X) = \one\{\E_0[Y|X] \geq \theta\}$ for the chosen threshold $\theta$. Figure \ref{fig:toy_example} (b) shows the mean difference in outcomes $\Delta$ from time 0 to time 1 as a function of $\theta$. (See the Appendix for a derivation.) This difference is around $1/3$, regardless of the $\theta$ chosen, indicating that more patients die as a result of the RAI. (In this scenario, $s(X)$ is bounded in [0, 0.30], so we only show thresholds in this range.)

The reason that all values of $\theta$ lead to worse outcomes is that $\theta$ corresponds to a threshold for $s(X) = \E_0[Y|X]$ rather than a threshold for $X$. In Figure \ref{fig:toy_example} (c) and (d), the vertical purple block represents the optimal treatment group $\{X \geq 0.22\}$, while the overlapping horizontal green block represents the group $\{\E_0[Y|X] \geq \theta\}$ that is actually treated under $d(X)$. Panel (c) shows the effect of choosing $\theta = 0.22$, the optimal threshold for $X$: we would fail to provide treatment to the group $\{X \geq 0.67\}$, indicated in red. This happens to be the group with the highest values of $\mathbb{E}[Y^0|X]$, i.e., the worst outcomes under no treatment. Conversely, Figure (d) shows the results of selecting the cutoff such that all those who would receive treatment under $\dopt(X)$ also receive treatment under $d(X)$: we wrongly treat the group $\mathbb{E}[Y^1|X] > \mathbb{E}[Y^0|X]$, again indicated in red.

These same problems can obviously arise in more complex scenarios, for example when $U \neq \emptyset$, when $X$ is high dimensional, and when the relationship between $X$ and the outcome is complex. In particular, we identify three properties of $s(X)$ that can give rise to these and other problems.

\subsection{$s(X)$ doesn't map nicely to a quantity of interest like $\E[Y^0|X]$, $\E[Y^1|X]$, or $d^\text{opt}(X)$}
Even though it is designed to predict outcomes under a historical treatment decision process, the RAI could usefully inform a new decision process if it bore some readily apprehensible relationship with a potential outcome-based quantity of interest. For example, if $s(X)$ were monotonic in $d^\text{opt}(X)$, then doctors might be able to use $s(X)$ to get closer to $d^\text{opt}(X)$, even without an explicit awareness of this relationship. In general, however, the relationship between $s(X)$ and any potential outcome-based quantity can be arbitrarily complex.

\subsection{Expertise can make things worse}
The more skilled doctors are at time $t=0$, the worse the system can get at time $t=1$. As an extreme example, if doctors are already behaving according to the optimal policy at time $t=0$, then necessarily, $\E_1[Y] \geq \E_0[Y]$. Alternatively, suppose that there are two medical systems $\P_0$ and $\P_0^*$ that are identical in their distribution of $(X, U, Y^0, Y^1)$. Imagine that they're also identical in terms of $A$, except that in system $\P_0^*$ doctors are more skilled at identifying who needs to be hospitalized:
\begin{align*}
	& \Pb_0^*\big(A = 1|\dopt(X, U) = 1\big) > \Pb_0\big(A = 1|\dopt(X, U) = 1\big)
\end{align*}
Then, under a threshold decision rule, we have that $\E_0^*[Y] < \E_0[Y]$ but $\E_1^*[Y] > \E_1[Y]$, so, perversely, people in system $\Pb_0^*$ are \textbf{better off} than people in system $\Pb$ at time 0 and \textbf{worse off} at time 1.

\subsection{The procedure is unstable under iteration}
Imagine that we iterate the process of gathering data from the system, developing a predictor, and implementing the threshold-based decision rule above. This seems like a plausible occurrence, in that as RAIs get integrated into more and more systems, necessarily any future data gathered from those systems will reflect the influence of those tools.

For time points $t = 1, 2, \ldots$, we have
\begin{align*}
	\E_t[Y|X] 
	              &= \one\{\E_{t-1}[Y|X] > \theta\}\E[Y^1|X] + (1 - \one\{\E_{t-1}[Y|X] > \theta\})\E[Y^0|X]
\end{align*}
Suppose we have some $X$ for which $\E_0[Y^1|X] < \theta, \E_0[Y^0|X] > t$ and $\E[Y_{(0)}|X] > \theta$. Then we'll have the situation depicted in Table \ref{t:iterated}, in which the treatment decision for this stratum just alternates at different time points. Ideally, as more and more data is gathered from a system, a decision procedure gets closer and closer to optimal. In this scenario, however, the treatment decision is the optimal one only at odd time points, while at even time points it's precisely the opposite.

\begin{table}
\centering
\begin{tabular}{l l l l}
Time $t$ & Treatment decision & $\E[Y_t|X]$ & $\E_t[Y|X]$ relative to $\theta$ \\
\hline
0 & Treat with probability $\pi_0(X)$ & $\E_0[Y|X]$ & $> \theta$ \\
1 & Treat all & $\E[Y^1|X]$ & $< \theta$ \\
2 & Treat none & $\E[Y^0|X]$ & $> \theta$ \\
3 & Treat all & $\E[Y^1|X]$ & $< \theta$ \\
4 & Treat none & $\E[Y^0|X]$ & $> \theta$ \\
$\ldots$ & & &
\end{tabular}
\caption{Treatment decisions and mean outcomes in stratum $X$ over time, under the deterministic decision rule that treats patients at time $t$ iff $\E_{t-1}[Y|X] > \theta$ for some threshold $\theta$, and assuming that $\E_0[Y^1|X] < \theta, \E_0[Y^0|X] > t$ and $\E[Y_{(0)}|X] > \theta$. \label{t:iterated}}
\end{table}

\section{Conclusion}
Decision makers choosing among different courses of action are naturally interested in the risk associated with each option. RAIs are in widespread use in many domains, but they are typically designed to predict outcomes under the historical decision process that generated the training data, rather than predicting potential outcomes under the available courses of action. This makes them generally unsuitable for informing future treatment or intervention decisions that are designed to reduce risk. Although previous work has proposed using potential outcome-based predictors in certain contexts, there has been little formal modeling of the consequences of current practice. Here, we show how RAIs based on observable rather than potential outcomes can plausibly lead to worse outcomes overall or for specific demographic groups than before their introduction, making them potentially both dangerous and unfair.

\bibliographystyle{unsrt}
\bibliography{bibliography.bib}

\appendix
\section{Derivations}
\subsection{Equation \eqref{eq:basic}: difference in mean outcome from time 0 to time 1}
Recall that we define
\begin{align*}
    & \pi_t(X, U) = \E_t[A|X, U] \\
    & \mu^a(X, U) = \E[Y^a|X, U]
\end{align*}
for $t = 1, 2, \ldots$ and $a \in \{0, 1\}$. Recall also that, per assumption 4, the distribution of the covariates $(X, U)$ doesn't change over time, so functions of $(X, U)$ don't change either.

For any time point $t$, we have
\begin{align*}
	\E_t[Y] &= \E_t\left\{ \E_t[AY^1 + (1-A)Y^0|X, U] \right\} \\
	      &= \E_t\left\{ \E_t[A|X,U]\E_t[Y^1|X, U] + (1-\E_t[A|X,U])\E_t[Y^0|X, U] \right\} \\
	      &= \E\left\{ \E_t[A|X,U]\E[Y^1|X, U] + (1-\E_t[A|X,U])\E[Y^0|X, U] \right\} \\
	      &= \E\left\{ \pi_t(X, U)\mu^1(X, U) + (1-\pi_t(X, U))\mu^0(X, U) \right\}	      
\end{align*}
where the second equality follows because $A\ind Y^a|X, U$ and the third equality follows from assumption 4. With $\Gamma(X, U) := \pi_1(X, U) - \pi_0(X, U)$, we have:
\begin{align*}
	\E_1[Y] - \E_0[Y] &= \E\left\{\pi_1(X,U)-\pi_0(X,U))(\mu^1(X,U) - \mu^0(X,U))\right\} \\
	                  &= \E\left\{\Gamma(X, U)(\mu^1(X,U) - \mu^0(X,U))\right\}
\end{align*}

\subsection{Calculating $\Delta$ in the toy example}
We have $U = \emptyset, X \sim \text{Unif}(0, 1)$, $\pi_0(X) = \E[Y^0|X] = X$, $\E[Y^1|X] = (0.7 - X)^2$, and $\pi_1(X) = \one\{\E_0[Y|X] \geq \theta\}$ for some chosen $\theta$. Plugging these into the above yields
\begin{align*}
    \Delta(\theta) = \E_1[Y] - \E_0[Y] &= \int \big(\one\{\E_0[Y|X] \geq \theta\} - X\big)((0.7 - X)^2 - X)d\mathbb{P}(X) \\
    &= \int \big(\one\{\E_0[AY^1 + (1 - A)Y^0|X] \geq \theta\} - X\big)\big((0.7 - X)^2 - X\big)d\mathbb{P}(X) \\
    &= \int \big(\one\{\pi_0(X)\mu^1(X) + (1 - \pi_0(X)\big)\mu^0(X) \geq \theta\} - X)\big((0.7 - X)^2 - X\big)d\mathbb{P}(X) \\
    &= \int \big(\one\{X(0.7 - X)^2 + (1 - X)X \geq \theta\} - X\big)\big((0.7 - X)^2 - X\big)d\mathbb{P}(X)
\end{align*}
where the second equality follows from the consistency assumption, and the third equality follows from the no unmeasured confounding assumption and assumption 4. This yields the curve in Figure \ref{fig:toy_example}(b).

\end{document}